\documentclass[aps,prl, reprint,10pt, superscriptaddress]{revtex4-1}
\usepackage[latin1]{inputenc}
\usepackage{graphicx}
\usepackage{amsmath,amssymb,amstext,graphicx,bm,booktabs}
\newcommand{\CS}{\ensuremath{C_\mathrm{S}} }

\newcommand{\fE}{\ensuremath{f_\mathrm{E}} }
\newcommand{\fM}{\ensuremath{f_\mathrm{M}} }
\newcommand{\fd}{\ensuremath{f_\mathrm{d}} }

\newcommand{\Gin}{\ensuremath{G_{\text{in}}}}
\newcommand{\Gout}{\ensuremath{G_{\text{out}}}}

\newcommand{\dIdfd}{\ensuremath{\mathrm{d}I/\mathrm{d}f_\mathrm{d}} }

\newcommand{\Vout}{\ensuremath{V_{\text{out}}}}
\newcommand{\VS}{\ensuremath{V_{\text{S}}}}

\newcommand{\VG}{\ensuremath{V_{\text{G}}}}

\newcommand{\St}{|\ensuremath{S}_{31}|}

\newcommand{\td}{\ensuremath{t_{\text{d}}}}
\newcommand{\tw}{\ensuremath{t_{\text{w}}}}
\newcommand{\tr}{\ensuremath{t_{\text{r}}}}

\newcommand{\tdelta}{\ensuremath{t_{\Delta}}}

\newcommand{\QE}{\ensuremath{Q_{\text{E}}}}
\newcommand{\QM}{\ensuremath{Q_{\text{M}}}}

\newcommand{\Vsd}{\ensuremath{V_{\text{sd}}}}

\newcommand{\go}{\ensuremath{g_0} }

\newcommand{\uQL}{\ensuremath{u_\mathrm{QL}}}

\newcommand{\CG}{\ensuremath{C_\mathrm{G}} }
\newcommand{\CCNT}{\ensuremath{C_\mathrm{CNT}} }

\newcommand{\VGbar}{\ensuremath{\overline{V_{\text{G}}}}}
\newcommand{\VTBbar}{\ensuremath{\overline{V_{\text{TB}}}}}

\newcommand{\Ztrans}{\ensuremath{Z_\mathrm{trans} }}

\newcommand{\uzp}{\ensuremath{u_\mathrm{ZP}}}
 
\newcommand{\Vzp}{\ensuremath{V_\mathrm{ZP}}}

\begin{document}

\title{Resonant optomechanics with a vibrating carbon nanotube and a radio-frequency cavity}

\author{N.~Ares}
\affiliation{Department of Materials, University of Oxford, Parks Road, Oxford OX1 3PH, United Kingdom}

\author{T.~Pei}
\affiliation{Department of Materials, University of Oxford, Parks Road, Oxford OX1 3PH, United Kingdom}

\author{A.~Mavalankar}
\affiliation{Department of Materials, University of Oxford, Parks Road, Oxford OX1 3PH, United Kingdom}

\author{M.~Mergenthaler}
\affiliation{Department of Materials, University of Oxford, Parks Road, Oxford OX1 3PH, United Kingdom}

\author{J.H.~Warner}
\affiliation{Department of Materials, University of Oxford, Parks Road, Oxford OX1 3PH, United Kingdom}

\author{G.A.D.~Briggs}
\affiliation{Department of Materials, University of Oxford, Parks Road, Oxford OX1 3PH, United Kingdom}

\author{E.A.~Laird}
\affiliation{Department of Materials, University of Oxford, Parks Road, Oxford OX1 3PH, United Kingdom}

\begin{abstract}

In an optomechanical setup, the coupling between cavity and resonator can be increased by tuning them to the same frequency. We study this interaction between a carbon nanotube resonator and a radio-frequency tank circuit acting as a cavity. In this resonant regime, the vacuum optomechanical coupling is enhanced by the DC voltage coupling the cavity and the mechanical resonator. Using the cavity to detect the nanotube's motion, we observe and simulate interference between mechanical and electrical oscillations. We measure the mechanical ring-down and show that further improvements to the system could enable measurement of mechanical motion at the quantum limit.

\end{abstract}

\date{\today{}}
\maketitle

The field of cavity optomechanics exploits resonant enhancement of light-matter interaction to study mechanical motion with exquisite sensitivity, including in the quantum domain~\cite{Aspelmeyer2014}. Commonly the mechanical frequency $\fM$ is much smaller than the frequency $\fE$ of the electromagnetic cavity, leading to a detuned coupling. The small zero-point motion then gives a comparatively weak optomechanical coupling, which however can be enhanced by a factor $\sqrt{n_c}$ through strong driving, where $n_c$ is the cavity photon occupation~\cite{Groblacher2009}. An alternative approach, which avoids the degradation of the cavity performance at low-temperatures for large $n_c$~\cite{Regal2008}, is to operate close to resonance, at $\fM \approx \fE$, with a DC voltage $\VGbar$ applied between the resonator and the cavity~\cite{Truitt2007, Bagci2014}. In this case, the single-photon coupling is enhanced by a factor $\sim \VGbar/V_\mathrm{ZP}$, where $V_\mathrm{ZP}$ is the zero-point fluctuation of the cavity voltage~\cite{Supp}. However, for this scheme to approach the quantum regime, it requires a mechanical resonator with a frequency high enough to be thermalized close to its ground state.

Here we realize a resonant optomechanical circuit exploiting the high mechanical frequency of a vibrating carbon nanotube.  Using a radio-frequency (RF) circuit as a readout cavity~\cite{Sillanpaa2009,Song2012}, we detect nanotube motion via its changing capacitance and measure interference between electrical and mechanical resonances. We show that the cavity measurement reproduces the resonance observed in transport, but go beyond transport by measuring the ringdown even without current through the nanotube. We reproduce the behavior in simulations, and show that feasible improvements to detection would approach the quantum limit.

Vibrating nanotubes offer attractive properties including high quality factors, large quantum zero-point motion, and high frequencies such that dilution refrigeration approaches the zero-phonon limit~\cite{Sazonova2004,Peng2006,Witkamp2006,Lassagne2008,Huttel2009,Laird2012,Schneider2014, Benyamini2014, Moser2014}. This has allowed development of force and mass sensors~\cite{Jensen2008,Chiu2008, Moser2013}, as well as proposals for coupling to electron spins~\cite{Ohm2012, Palyi2012} and optical photons~\cite{Rips2012,Rips2013}. A fiber cavity was recently used to detect room-temperature Brownian motion~\cite{Stapfner2013}, but cryogenic measurements to date rely on electrical transport through the nanotube, forgoing potential for sensitivity and coherent control offered by coupling to an chip-scale cavity.

\begin{figure}
\includegraphics{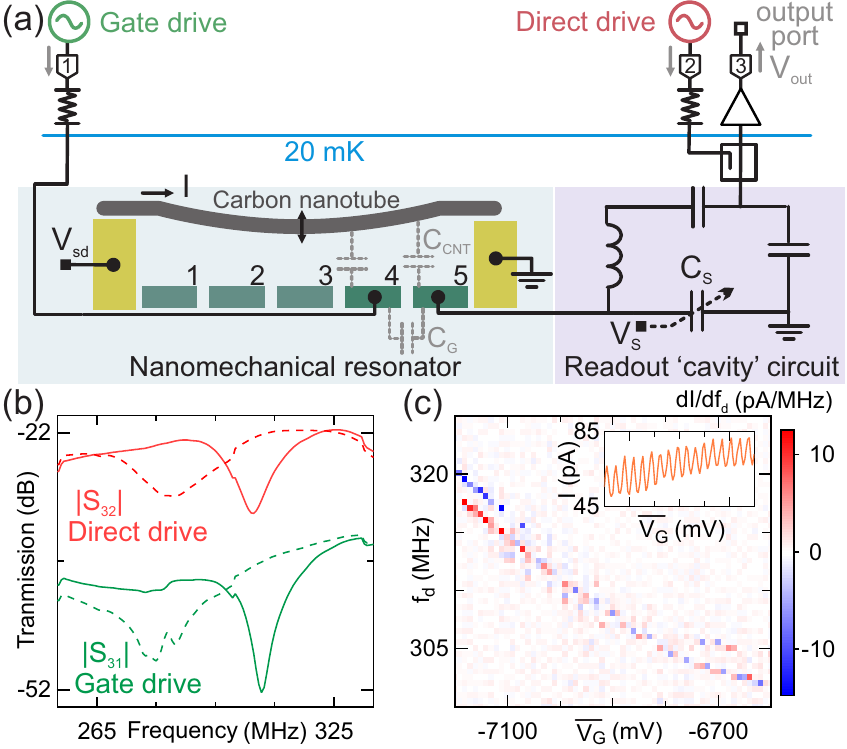}
\caption{\label{Fig1} (a)  Experimental setup. A gated and contacted carbon nanotube resonator is connected to a radio-frequency tank circuit acting as a readout cavity. Nanotube motion is excited via a gate drive; the cavity can independently be probed directly via a directional coupler. The cavity response is detected via a cryogenic amplifier. The combined setup acts as a three-port RF device, with transmission measurements possible from ports 1 or 2 to port 3. Dotted capacitors indicate coupling between gates 4 and 5 and between these gates and the nanotube. DC voltages $\VGbar_1$ to $\VGbar_3$ are applied to gates 1-3. Bias tees (not shown) allow DC voltages $\VGbar_4$ and $\VGbar_5$ to be added to the RF signal on gates 4 and 5, respectively. (b) Transmission as a function of frequency with cavity tuning voltage $\VS=9$~V (solid line) and $\VS=0.2$~V (dotted line) for both direct and gate drive, showing a tuneable resonance. The different insertion losses are mainly due to different fixed attenuators in the two paths. (c) Numerically differentiated $\dIdfd$ as a function of $\fd$ and $\VGbar$. The gate drive power was P$=-33$~dBm. Inset shows $I$ over the same $\VGbar$ range with $\Vsd=10$~mV and with no RF excitation. 
 }
\end{figure}

A carbon nanotube resonator [Fig.\ref{Fig1}(a)] was suspended between contact electrodes over an array of five finger gates as described in Ref.~\cite{Supp, Mavalankar2016}. For transport measurements the device is probed by applying a bias $\Vsd$ and measuring the current $I$ through the nanotube. The doping of the nanotube, and hence its conductance, are tuned using DC gate voltages $\VGbar_1$ to $\VGbar_5$; these simultaneously tune the mechanical tension~\cite{Sazonova2004}. We define $\VGbar$ as the DC voltage applied to all gate electrodes. Gate 4 is connected to an RF source that can be used to excite transverse vibrations. Measurements are at 13~mK in a dilution refrigerator.

The effective RF cavity is realized using an inductor and capacitors mounted on the sample holder and coupled to gate 5. To tune cavity frequency $\fE$, we incorporate a variable capacitor $\CS$ controlled by a DC voltage $\VS$, giving $\fE \approx (2\pi\sqrt{L\CS})^{-1}$, where $L=180$~nH is the circuit inductance~\cite{Ares2016}. 
This cavity can be driven in two ways; by injecting an RF signal to the input via a directional coupler (direct drive), and by driving via gate 4, which is capacitively coupled (gate drive). The cavity output is fed to a cryogenic amplifier and detected at room temperature. The entire setup forms a three-terminal circuit with input ports 1 and 2 and output port~3.  Figure~\ref{Fig1}(b) shows electrical scattering parameters as a function of driving frequency $\fd$ for two settings of $\VS$ in both direct and gate drive. The cavity resonance is evident as a transmission minimum with quality factor $\QE \approx 10-20$ depending on varactor losses~\cite{Ares2016}.

The mechanical resonance is first studied in transport [Fig.\ref{Fig1}(c)]~\cite{Sazonova2004}. A hole-doped quantum dot potential is created through a combination of Schottky barriers at the contacts and a voltage $\VGbar$. As a function of $\VGbar$, a series of Coulomb peaks is evident in transport [Fig.\ref{Fig1}(c) inset]. The effect of mechanical displacement is to change the capacitance between the dot and the gates~\cite{Huttel2009}, shifting the Coulomb peaks in a way that leads to a rectified DC current $I$. With an RF signal applied to gate 4, this is evident when $\dIdfd$ is plotted as a function of $\VGbar$ and $\fd$~\cite{Huttel2009, Laird2012, Benyamini2014}. A resonance feature is seen at the mechanical frequency $\fM$ [Fig.~\ref{Fig1}(c)]. The measured $\fM$ is consistent with similar nanotube resonators~\cite{Sazonova2004,Peng2006,Witkamp2006,Chiu2008,Lassagne2008,Huttel2009, Schneider2014}. The mechanical nature is clear from the dependence on $\VGbar$ due to electrostatic tensioning.

\begin{figure}
\includegraphics[width=\columnwidth]{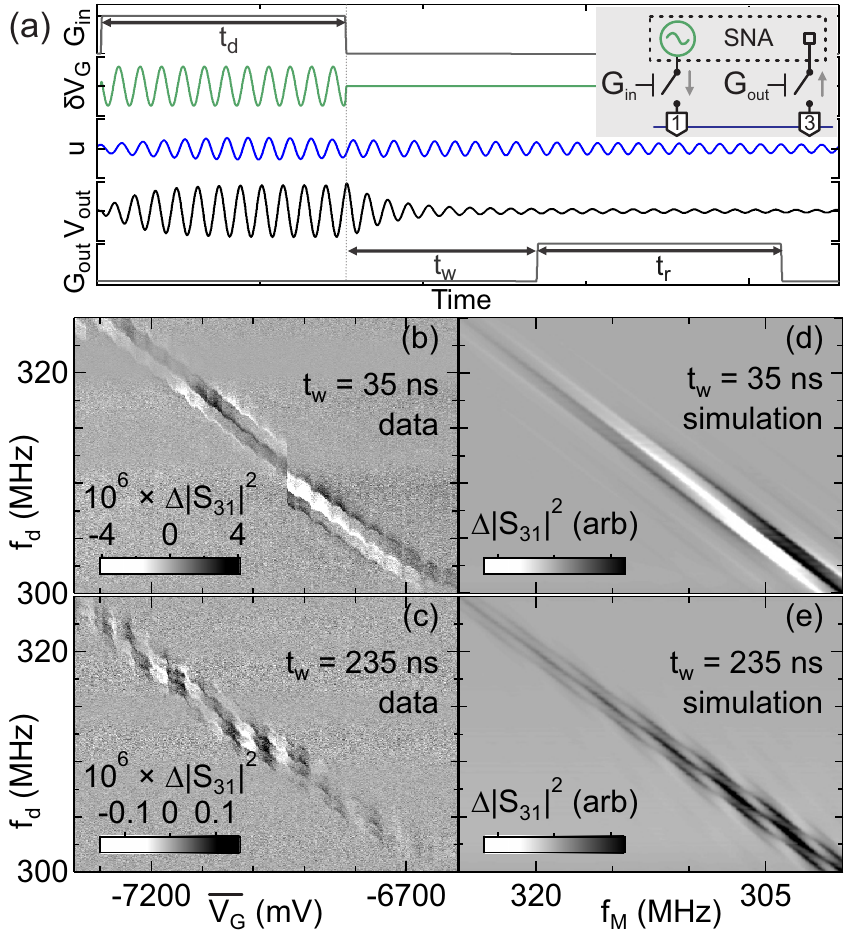}
\caption{(a) Schematic pulsed excitation and measurement scheme, showing gating signals $\Gin$ and $\Gout$ together with simulated drive $\delta \VG$, displacement $u$, and cavity output voltage $\Vout$. Durations of drive ($\td$), wait ($\tw$) and acquisition ($\tr$) are indicated. Inset: room-temperature gating scheme using RF switches.
 (b), (c) Time-averaged $\Delta\St^2$, measured using the scheme in (a), as a function of $\VGbar$ and $\fd$ for two different values of $\tw$. Each column of data is averaged for 5~s. Measurement parameters were $\VS=9$~V, $\Vsd=2$~mV, $P=-38$~dBm and $\td=\tr=300$~ns. The jump at $\VGbar \approx \-6900$~mV indicates an electrostatic switcher in the device. (d), (e) Simulated $\Delta\St^2$ (see text) reproducing panels (b) and (c) respectively. 
\label{Fig2}
}
\end{figure}

We now detect this motion via the readout cavity. The dependence of gate capacitance $C_\mathrm{CNT}$  on mechanical displacement $u$ leads to a cavity coupling $\frac{d \fE}{d u} \approx \frac{\fE}{2\CS}\frac{\partial \CCNT}{\partial u}$~\cite{Supp}. In a detuned optomechanical setup with $\fM \ll \fE$, this coupling allows the motion to be monitored via the phase shift of the cavity transmission~\cite{Regal2008, Aspelmeyer2014}. We use a different scheme that takes advantage of the fact that $\fM$ and $\fE$ can be tuned into resonance. With the DC voltage on gate 5 set to $\VGbar$, the mechanical motion induces a current $I_\mathrm{G} = -\VGbar \frac{\partial \CCNT}{\partial u} \frac{du}{dt}$ onto the gate electrode. This current excites the cavity, leading to an output signal 
\begin{equation}
\Vout = \Ztrans I_\mathrm{G},
\end{equation}
where $\Ztrans$ is a transimpedance set by the circuit parameters~\cite{Supp}. The transimpedance, and hence the sensitivity, is maximal close to degeneracy ($\fM \approx \fE$). In contrast to detuned readout (at least outside the resolved-sideband regime), the signal arises from velocity $\frac{du}{dt}$ rather than displacement.

We demonstrate this coupling with the cavity tuned to $\fE =304$~MHz and with $\VGbar = -7$~V. Drive and detection are  provided by a scalar network analyser (SNA), which monitors the time-average transmission $\St$ from port~3 to port 1. A challenge of this scheme is that the drive used to actuate motion also couples directly to the cavity via a parasitic capacitance $\CG$ between gates 4 and 5. This gives rise to a non-resonant background that in our unoptimized circuit overwhelms the mechanical signal. (This is why a mechanical signal is not seen directly in Fig.~1(b)).
We can largely reject this contribution using the pulsed driving scheme of Fig.~2(a), which separates the times of driving and detection using RF switches to gate the input and output signals. With the output decoupled, gate 4 is first driven for time $\td = 300$~ns, exciting both the mechanical motion and the cavity voltage. After a wait interval $\tw$ from the end of this drive burst, the output is coupled for a detection interval of duration $\tr=300$~ns. For each data point, the  detected power was averaged over $\sim 8$~ms during which this cycle was applied with a period of $\tw+610$~ns. By choosing $\tw$ intermediate between the cavity and mechanical ringdown times ($\QE/2\pi\fE \approx 11~\mathrm{ns}>\tw >\QM/2\pi\fM \approx 370$~ns), the mechanical contribution to $\Vout$ is largely isolated.

Figure 2(b-c) shows transmission as a function of $\fd$ and $\VGbar$ for two different values of $\tw$. To highlight the signal due to the nanotube, data are plotted after subtraction of the average $\St^2$ at each frequency, giving the excess signal $\Delta \St^2$. The mechanical resonance is evident, with a similar dependence of $\fM$ on $\VGbar$ as measured in Fig.~1(c). However, it now appears as a pattern of bright and dark fringes whose alignment depends on $\tw$. This is attributable to a classical interference effect between the ringdown of mechanical and electrical signals. Both the nanotube and the cavity are set ringing during the drive burst with phases that depend on the difference between their corresponding resonance frequencies and $\fd$. Subsequently each evolves at its own frequency. The two signals therefore add constructively or destructively in a way that depends on $\tw$, $\fd$, $\fE$, and $\fM$ (tuned by $\VGbar$). Superimposed on this pattern are sidebands separated by the gating repetition frequency. The data also show a stairlike dependence of $\fM$ on $\VGbar$ because Coulomb blockade modulates the electrostatic tensioning~\cite{Sapmaz2003, Steele2009, Lassagne2009}.

Both fringes and sidebands are reproduced in a simulation~\cite{Supp} of $\Vout(\fM, \fd)$ by modelling the electrical and mechanical impedances between gate 4 and the cavity output~[Fig.\ref{Fig1}(c-d)]. From the sharpness of the resonance and the intensity of the interference fringes, the mechanical quality factor can be estimated as $\QM \approx 700$, while the effective coupling is $\partial \CCNT / \partial u \sim 0.3\times10^{-12}~\mathrm{Fm}^{-2}$, roughly consistent with the device geometry~\cite{Supp}. Coulomb blockade effects are not seen in the simulation because results are plotted as a function of $\fM$ rather than $\VGbar$.

\begin{figure}
\includegraphics[width=\columnwidth]{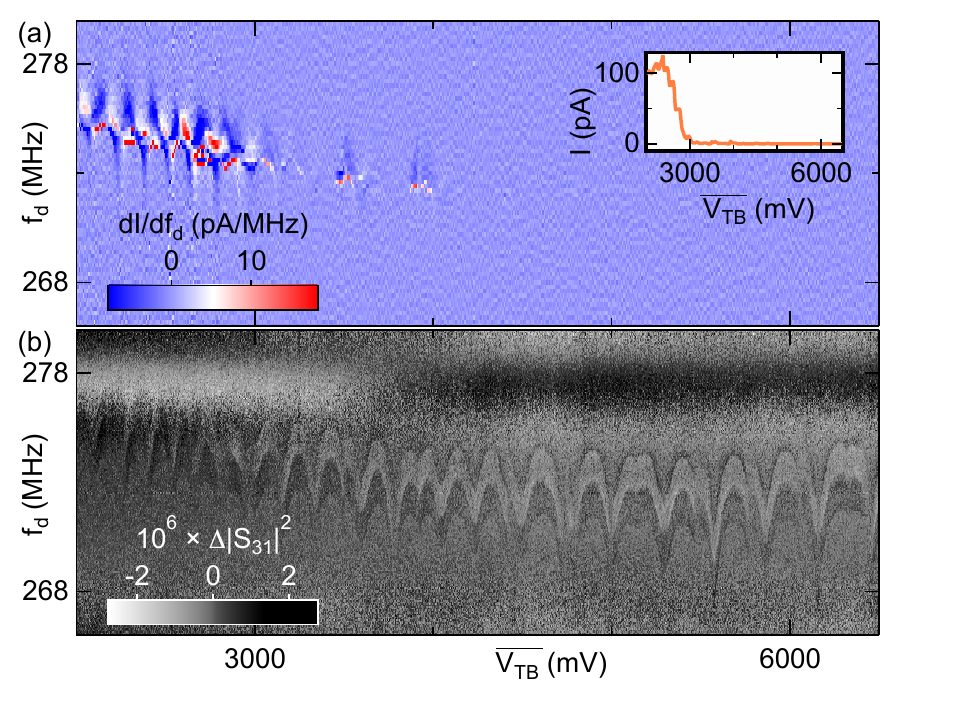}
\caption{
(a) Numerically differentiated $\dIdfd$ as a function of $\fd$ and $\VTBbar$ for $P=-38$~dBm and $\Vsd=10$~mV. The inset shows $I$ for the same range of $\VTBbar$ and same $\Vsd$ with no microwaves applied. As $\VTBbar$ increases, the tunnel barriers are pinched off so that $I$ cannot be detected. 
(b) Measured $\Delta\St^2$ following the scheme in Fig.~\ref{Fig2}(a) and plotted as a function of $\fd$ and $\VTBbar$. The resonance is seen even with transport pinched off. Measurement parameters were $\VS=0.2$~V,  $\td=\tr=300$~ns and $\tw\sim15$~ns.
\label{Fig3}
}
\end{figure}

In contrast to previous electromechanical measurements of nanotubes~\cite{Sazonova2004,Peng2006,Chiu2008,Witkamp2006,Lassagne2008,Huttel2009,Laird2012,
Schneider2014}, cavity readout allows detection of mechanical motion even for device configurations where $I$ is unmeasurably small. To access such a configuration, we set $\VGbar_2 = \VGbar_3 =\VGbar_4= -7250$~mV and we adjust $\VGbar_1 = \VGbar_5\equiv \VTBbar$ to tune the dot tunnel barriers. Figure~3 inset shows how, as an increased $\VTBbar$ makes tunnel rates smaller, transport turns-off to an undetectable level ($I \ll 1$~pA) when $\VTBbar \gtrsim 4000$~mV. As expected, the mechanical resonance becomes correspondingly unmeasurable in transport [Fig.~3(a)]. However, it is evident in cavity readout across the entire range [Fig. 3(b)]. The oscillations of $\fM$ as a function of $\VTBbar$ are an effect of mechanical softening on Coulomb blockade peaks~\cite{Steele2009, Lassagne2009} and indicate that the sum of the dot's tunnel rates to left and right, although not big enough to lead to a detectable current, is at least of the order of $\fM$ ($\Gamma_\mathrm{L} + \Gamma_\mathrm{R} \gtrsim \fM$).
  
\begin{figure}
\includegraphics[width=\columnwidth]{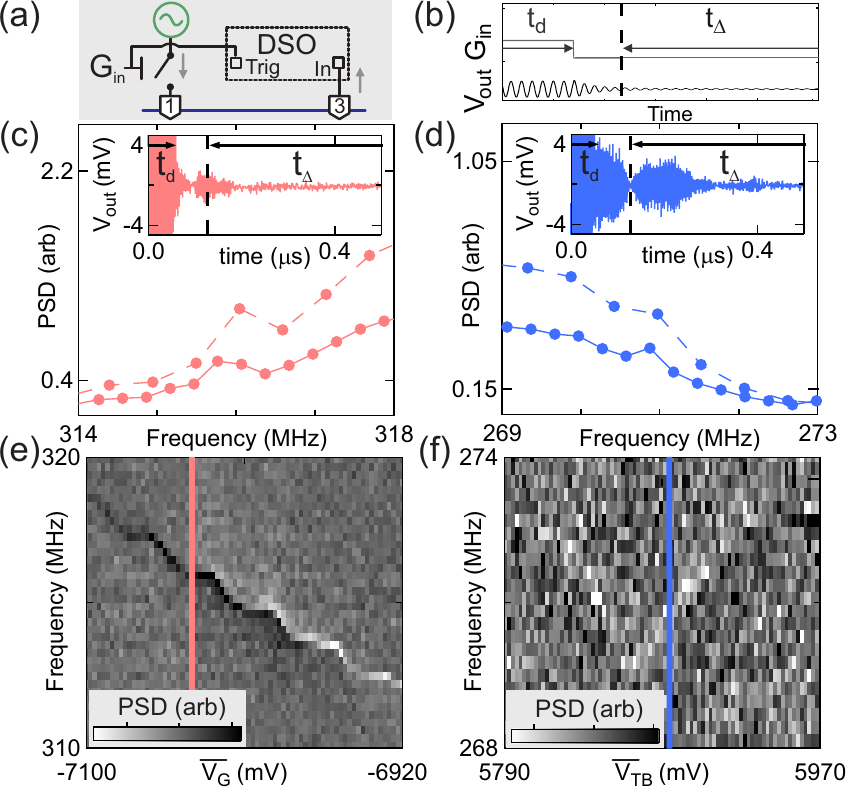}
\caption{\label{Fig4}
(a) Gating scheme using a RF switch to gate the drive and a digital sampling oscilloscope (DSO), triggered by the gating signal $\Gin$, to capture $\Vout$. (b) Schematic of $\Gin$ and $\Vout$. Durations of drive ($\td$) and acquisition ($\tdelta$) are indicated. 
(c), (d)  Averaged PSD for two fixed gate voltages marked in (e), (f), respectively. Solid (dashed) curves correspond to $\tdelta=3.32~\mu$s ($\tdelta=1.82~\mu$s). Measurement parameters were $\Vsd=10$~mV, $P=-38$~dBm, $\td=330$~ns and time between drive pulses was $3660$~ns. For (c), $\fd=315$~MHz and $\VS=9$~V, while for (d), $\fd=273$~MHz and $\VS=0.2$~V. 
Insets show averaged $\Vout$ traces. The drive burst ends $\sim50$~ns after the beginning of these traces and acquisition begins $\sim80$~ns  later (dashed lines). 
(e), (f) Measured PSD as a function of frequency and gate voltage for the two different device configurations studied, averaged over 65500 repetitions at each gate voltage. A background due to purely electrical resonances was subtracted. 
The signal-to-noise ratio is better in (e) than in (f), presumably reflecting larger electrical coupling when the nanotube is in a more conducting state.
}
\end{figure}

We now study the mechanical decay rate for different tunnel barrier configurations by directly measuring the ringdown. A time-resolved measurement was previously performed using a fast current amplifier~\cite{Schneider2014}, but relied on transport through the device. Since for some barrier configurations, $\QM$ is known to be limited by tunneling resistance~\cite{Steele2009, Lassagne2009, Benyamini2014, Moser2014}, it is of interest to determine the limiting $\QM$ in an electrically pinched-off nanotube. Extracting a precise linewidth from data as in Fig.~2 is complicated by the interference fringes. We therefore sample the cavity output directly in the time domain. For this experiment, port 1 is connected to an RF source generating a fixed frequency $\fd$ gated with a function $\Gin(t)$ and port 3 is connected without gating to an oscilloscope that digitizes $\Vout$ [Fig.~\ref{Fig4} (a-b)]. Typical traces averaged over $\sim 3500$ repetitions are shown in  Fig.~\ref{Fig4} (c-d) insets. Because the electrical ringdown is close in frequency to the mechanical one, it is not straightforward to separate them in the time domain. Instead, we take a segment of duration $\tdelta$ from each averaged trace and transform it to give a frequency power spectral density (PSD), in which the mechanical signal appears as a separate peak. For acceptable signal-to-noise, we further average PSD over 1250 traces [Figs.~\ref{Fig4} (c-d)].

Figure~\ref{Fig4} (e-f) shows two-dimensional maps of PSD plotted as a function of frequency and gate voltage in two gate voltage ranges; one `open' [part of the range in Fig.~\ref{Fig2}(b-c)] and one `closed' [part of the range in Fig.~3]. After subtracting a background due to purely electrical resonances, the mechanical signal is evident as a gate-dependent peak or dip in PSD. 

We calculate $\QM$ from the linewidth in this PSD. Figure 4(c-d) shows spectral densities at fixed gate voltage settings in open and closed configurations. The mechanical signal is evident as a weak but sharp peak superimposed on a broad background arising from several electrical resonances of the cryostat. The choice of $t_\Delta$ is set by a tradeoff; larger $t_\Delta$ increases the frequency resolution in the PSD and suppresses the background contribution due to the cavity, but also weakens the mechanical peak by incorporating more of the decay. This tradeoff is illustrated by traces for two choices of $t_\Delta$ [Fig.~4 (c-d)]. For an open gate configuration [Fig.~4(c)] the mechanical peak is resolved to a width of $\Delta \fM\approx0.3~$MHz for $t_\Delta = 3.32~\mu$s, allowing the estimate $\QM = \fM/ \Delta \fM \approx 1000$. For the closed gate configuration [Fig.~4(d)] the peak's width is not resolved and only a lower bound $\QM \gtrsim 900$ is extracted. In this device there is therefore no evidence that $\QM$ is limited by tunnel resistance, despite the modest value compared with other clean nanotube devices~\cite{Steele2009, Huttel2009, Moser2014}. Since the sensitivity does not yet allow single-shot readout, this is a lower bound, incorporating dephasing as well as decoherence~\cite{Schneider2014}. Interestingly, $\QM$ is slightly higher than measured previously for a stamped nanotube~\cite{Benyamini2014}.

To set this work in context with other implementations~\cite{Aspelmeyer2014}, we calculate several measures of the optomechanical interaction, using the geometrically estimated coupling capacitance~\cite{Supp}. The vacuum optomechanical coupling $\go = 2\pi\fE \frac{\partial \fE}{\partial u} u_\mathrm{zp}$, which is the cavity shift due to zero-point displacement $u_\mathrm{zp} = \sqrt{\hbar/4\pi m\fM} \approx 2$~pm, is $\go\approx2\pi\times0.4$~mHz, where $m$ is the nanotube mass~\cite{Supp, Poot2012}. For typical cavity drive of -38~dBm, we estimate the photon occupation as $n_c \approx 8\times10^{9}$, so that the detuned optomechanical strength is $g_D = \sqrt{n_c} \go \approx 2\pi\times35$~Hz. However, the cavity-nanotube interaction near $\fM=\fE$ is parameterized by a resonant coupling $g_R=\frac{\VGbar}{\Vzp} \go$, where $\Vzp \sim 240$~nV is the zero-point voltage on the cavity electrode. Our parameters imply $g_R \approx 2\pi\times 12$~kHz, so that this coupling dominates~\cite{Supp}.

This resonant enhancement might allow quantum-limited measurements in the ground state. In a continuous displacement measurement, the uncertainty principle~\cite{Caves1982, Braginsky1995, LaHaye2004} limits the position resolution to $\Delta \uQL = \sqrt{2/\ln3}\,\uzp$. In a detuned configuration, where measurement is via phase shift to a cavity probe tone, the attainable imprecision at low temperature is typically limited by cavity nonlinearity at high drive power~\cite{Regal2008}. In our setup, the analog of the probe tone is $\VGbar$, which does not directly excite the cavity. Given an amplifier voltage sensitivity~$S_V$, Eq.~(1) predicts a vibrational amplitude sensitivity $S_u = S_V/(2\pi\fM\Ztrans\VGbar\frac{\partial \CCNT}{\partial u})$, which for our unoptimized parameters is $S_u \approx 6~\mathrm{pm} / \sqrt{\mathrm{Hz}}$~\cite{Supp}. The corresponding imprecision is $\Delta u = S_u \sqrt{\Delta \fM} \approx 1700\times\Delta \uQL$, limited by amplifier noise. 
To resolve $\Delta \uQL$ would require a near-quantum-limited external amplifier such as a SQUID~\cite{Clarke2006, Clerk2010} as well as an increase in $\Ztrans$, $\QM$, or the coupling strength. With other parameters unchanged, increasing gate length to 550 nm and $\VGbar$ to 65~V would give a quantum backaction imprecision of order $\Delta \uQL$~\cite{Supp}. This would allow quantum effects to be studied in a resonator near the phonon ground state~\cite{Supp,Teufel2009, Anetsberger2010}.

\begin{acknowledgments}
We acknowledge discussions with E.M.~Gauger and support from EPSRC (EP/J015067/1), DSTL, Marie Curie CIG and IEF fellowships, the Royal Academy of Engineering, and Templeton World Charity Foundation. The opinions expressed in this publication are those of the authors and do not necessarily reflect the views of Templeton World Charity Foundation. NA performed the experiment. TP developed the fabrication process and contributed to equipment setup. AM fabricated and characterised the device and built the tank circuit. JHW set up the CVD furnace. EAL and NA conceived the experiment and analyzed the data. All authors discussed results and commented on the paper.
\end{acknowledgments}

\vfill

\end{document}


\renewcommand{\theequation}{S\arabic{equation}}
\renewcommand{\thefigure}{S\arabic{figure}}
\setcounter{secnumdepth}{3}
\setcounter{tocdepth}{3}
\onecolumngrid
\parbox[c]{\textwidth}{\protect \centering \Large \MakeUppercase{Supplementary Material}}
\rule{\textwidth}{1pt}
\vspace{1cm}
\twocolumngrid

\section{Device fabrication}

To fabricate the nanotube, FeCl$_3$ catalyst was mixed with PMMA and spun on a quartz substrate previously etched with pillars. Nanotubes were synthesized on this substrate by chemical vapour deposition  at 950~$^\circ$C from a CH$_4$/H$_2$ mixture diluted to $20~\%$ concentration in argon. Following growth, this quartz chip was aligned with the device chip in an optical mask aligner and nanotubes were transferred by stamping.
The diameter of the particular carbon nanotube used in this experiment was not measured.  

\section{Detuned and resonant optomechanics}

This section illustrates the equivalence of detuned and resonant optomechanical Hamiltonians and shows how the coupling parameters are related. The basic optomechanical setup is shown in Fig.~\ref{FigS0}. 
We begin with the Hamiltonian of the cavity, the resonator, and the displacement-dependent capacitor $\CCNT(u)$ connecting them:
\begin{equation}
H = \hbar \omega_E \ahat^\dagger \ahat + \hbar \omega_M \bhat^\dagger \bhat + \frac{1}{2}\CCNT V^2
\label{eq:Hamiltonian}
\end{equation}
where $\omega_E = 2\pi \fE$ and $\omega_M = 2\pi \fM$ are the cavity and mechanical angular frequencies, $\ahat$ and $\bhat$ are photon and phonon annihilation operators, and $u$ is mechanical displacement. Optomechanical coupling arises through the displacement dependence of $\CCNT$, which enters directly in the last term and by modifying $\omega_E$ in the first term; the dependence of $\omega_E$ on $u$ is given by:
\begin{equation}
\omega_E \approx \omega_E^0 - \frac{\go}{\uzp}u
\end{equation}
where
\begin{equation}
g_0 \approx \frac{\omega_E^0}{2\CS} \frac{\partial \CCNT}{\partial u} \uzp
\label{eq:g0}
\end{equation}
is the single-photon coupling strength. Here $\omega_E^0$ is the cavity frequency for zero displacement, $\CS$ is the total cavity capacitance, and $\uzp$ is the zero-point motion.

\subsection{Detuned optomechanical coupling}
\begin{figure}
\includegraphics[width=\columnwidth]{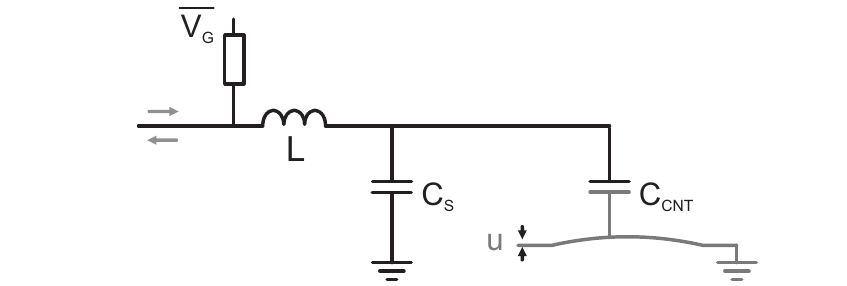}
\caption{\label{FigS0}
Simplified optomechanical setup. The cavity is modelled as an $LC$ resonator with effective inductance $L$ and capacitance $\CS + \CCNT$, with $\CCNT \ll \CS$. Coupling to the resonator's displacement $u$ is via the displacement dependence of the coupling capacitor $\CCNT$. The coupling capacitor is biased by DC voltage $\VGbar$ via a large resistor as shown, which gives rise to a resonant coupling as discussed in the Supplementary text.}
\end{figure}

If, as in many optomechanical implementations, no DC voltage is applied, the last term of Eq.~(\ref{eq:Hamiltonian}) can be neglected as second order. In this situation, a linearized Hamiltonian can be derived~(see e.g. Ref.~\onlinecite{Aspelmeyer2014}) by transforming Eq.~(\ref{eq:Hamiltonian}) to a frame rotating at the cavity drive frequency and approximating $\ahat \approx \sqrt{n_c} + \delta \ahat$, where $n_c$ is the average cavity photon occupation. The linearized Hamiltonian is:
\begin{equation}
H^\mathrm{(lin)}= -\hbar \Delta \delta \ahat^\dagger \delta \ahat + \hbar \omega_M \bhat^\dagger \bhat - \hbar \gD (\delta \ahat^\dagger + \delta \ahat)(\bhat^\dagger + \bhat)
\label{eq:Hamiltoniandispersive}
\end{equation}
where $\Delta$ is the frequency detuning of the cavity drive and $\gD\equiv\sqrt{n_c}g_0$ is the optomechanical coupling strength. In the circuit of Fig.~\ref{FigS0} the photon number can be expressed in terms of the driven gate voltage by $n_c = V_\mathrm{AC}/\Vzp$, where $V_\mathrm{AC}$ is the amplitude of the driven gate voltage and $\Vzp$ is the zero-point amplitude. The optomechanical coupling is then
\begin{equation}
\gD=\frac{V_\mathrm{AC}}{\Vzp}g_0.
\end{equation}

The origin of the coupling in Eq.~(\ref{eq:Hamiltoniandispersive}) is the dependence of $\omega_E$ on $u$; it leads to energy exchange between mechanical and cavity degrees of freedom when $\Delta \approx \pm \omega_M$, which allows $\omega_M$ and $\omega_E$ to be very different.

\subsection{Resonant optomechanical coupling}
We now consider the situation where the third term of Eq.~(\ref{eq:Hamiltonian}) cannot be neglected. We expand it by writing:
\begin{eqnarray}
\CCNT 	&=& \overline{\CCNT} + \frac{\partial \CCNT}{\partial u} u \\
V		&=& \VGbar + \delta V.
\end{eqnarray}
Here $\CCNTbar$ represents the static part of the capacitance, $\VGbar$ is the DC gate voltage, and $\delta V$ is the AC cavity voltage. To second order in $u$ and $\delta V$, we therefore have
\begin{equation}
\CCNT V^2 = \CCNTbar(\VGbar+\delta V)^2 + \VGbar^2 \delta C + 2 \VGbar \delta V \frac{\partial \CCNT}{\partial u} u.
\label{eq:suppcoupling}
\end{equation}
The first term of Eq.~(\ref{eq:suppcoupling}) simply renormalizes $\omega_E$, while the second term shifts the equilibrium displacement. We therefore neglect these, leaving the third term representing an optomechanical coupling. Writing $u$ and $\delta V$ in terms of ladder operators and substituting into Eq.~(\ref{eq:Hamiltonian}) gives:
\begin{equation}
H = \hbar \omega_E \ahat^\dagger \ahat + \hbar \omega_M \bhat^\dagger \bhat + \hbar \gR (\ahat^\dagger + \ahat)(\bhat^\dagger + \bhat).
\label{eq:Hamiltonianresonant}
\end{equation}
The optomechanical coupling is now $\gR=\partial \CCNT / \partial u \cdot \VGbar \Vzp \uzp/ \hbar$, where $\Vzp$ is the zero-point voltage on the cavity electrode. We see that Eq.~(\ref{eq:Hamiltonianresonant}) is isomorphic to~Eq.~(\ref{eq:Hamiltoniandispersive}) with the substitutions $\omega_E \leftrightarrow \Delta$, $\ahat \leftrightarrow \delta \ahat$, and $g_D \leftrightarrow g_R$, confirming that the  physics of linear optomechanics is accessible in this system. However, the Hamiltonian~(\ref{eq:Hamiltonianresonant}) only allows energy exchange when $\omega_M \approx \omega_E$, implying that this is a resonant coupling.

\begin{figure}
\includegraphics[width=\columnwidth]{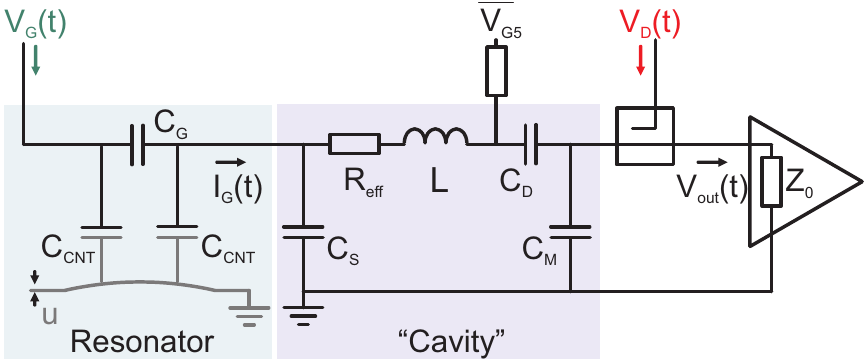}
\caption{\label{FigS1}
Schematic of the optomechanical circuit from Fig.~1. The tank circuit is formed by components $L$, $\CS$, $\CD$,  $\CM$, and an electromechanical capacitance $\CCNT$ depending on the nanotube displacement $u(t)$. The effective parasitic resistance $\Reff$ parameterizes circuit losses. The tank circuit is probed either in reflection mode (by applying voltage $\VD$ through a directional coupler) or in transmission mode by applying voltage $\VG$, which also excites the mechanical resonator. The output voltage $\Vout$ is monitored with an amplifier of impedance~$Z_0$.}
\end{figure}

To compare the coupling constants in the two regimes, we write $\Vzp=\sqrt{\hbar \omega^0_E/2\CS}$. Then the resonant coupling can be written
\begin{equation}
\gR=\frac{\VGbar}{\Vzp} \go,
\end{equation}
implying that the enhancement over the detuned case is:
\begin{equation}
\frac{\gR}{\gD}=\frac{\VGbar}{V_\mathrm{AC}}.
\end{equation}
For many types of device, including the one measured here, it is easier to establish a DC gate voltage $\VG$ than a comparable driven voltage $\VAC$. For example, in this experiment, $\VG \sim 7$~V while for -40~dBm of driving, limited by the cryostat cooling power, we estimate $\VAC \sim 30$~mV. If the resonant condition $\fE \approx \fM$ can be achieved, the potential for  enhancement of the optomechanical coupling is therefore substantial.

\section{Simulations for Fig.~2}

This section describes how the simulations for Fig.~2 were performed.

\begin{figure}
\includegraphics[width=\columnwidth]{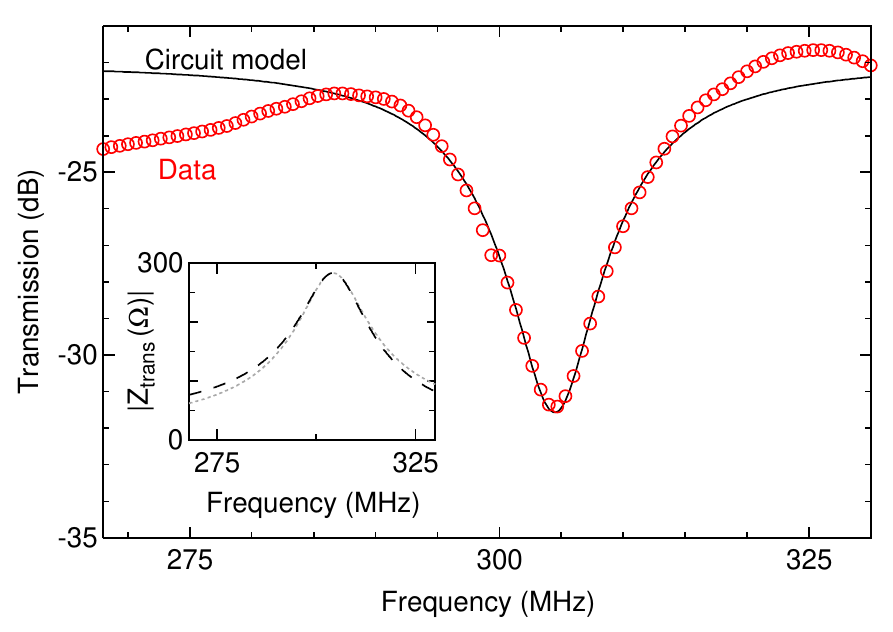}
\caption{\label{FigS2}
Symbols: Measured transmission $|S_{32}|$, proportional to the reflection from the tank circuit. Curve: simulation using circuit model of Fig.~\ref{FigS1} (see text for parameters). An overall offset is applied to the model to account for fixed insertion losses in the cryostat. Inset: Transimpedance simulated by Eq.~(\ref{eq:Ztrans}) using these circuit parameters (dashed line) and approximated by Eq.~(\ref{eq:Ztransapprox}) (dotted line).
}
\end{figure}

\subsection{The optomechanical circuit}
\label{sec:circuit}

Both the circuit and the mechanical resonator were simulated classically. Figure~\ref{FigS1} shows a schematic of the optomechanical circuit. The resonator, modelled as a grounded  conductor, is coupled via displacement-dependent capacitances $\CCNT(u)$ to the drive gate (gate~4 or G4) and the pickup gate (gate 5 or G5). As well as the coupling via the resonator, there is also a geometric capacitance $\CG \gg \CCNT$ between these two gates. The ``cavity'' tank circuit~\cite{Ares2016} is formed from an inductor $L$ and a combination of fixed and variable capacitors, with $\CS$ and $\CM$ incorporating varactor diodes tuned by DC bias voltages. Dissipation is parameterized by an effective resistance $\Reff$. The DC gate bias $\overline{V_\mathrm{G1}}$, which tunes the coupling to the resonator, is applied via a bias tee. As in Fig.~1, the reflection of the cavity can be probed using a signal $\VD(t)$ injected via a directional coupler.

\begin{figure*}
\includegraphics[width=17.8cm]{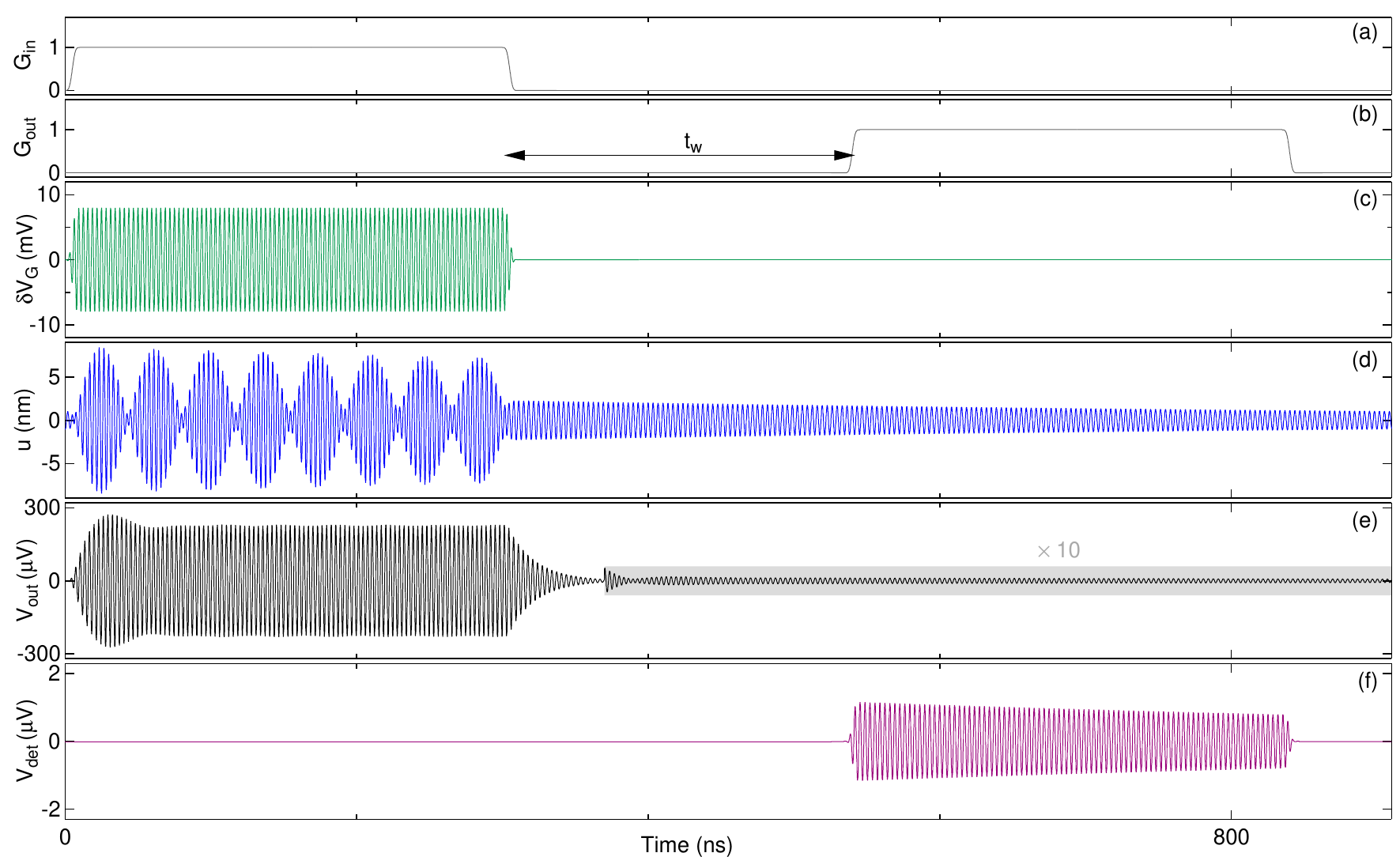}
\caption{\label{FigS3}
Simulated signals during one cycle of pulsed driving. (a,b) Gating functions $G_\mathrm{in}(t), G_\mathrm{out}(t)$ of drive and detection respectively; (c) Gate voltage driving signal $\delta \VG(t)$; (d) nanotube displacement $u(t)$; (e) cavity output $\Vout(t)$ (shaded area magnified tenfold); (f) gated output $\Vdet(t)$. The slow decay of the mechanical oscillation is clear in (d), compared with the rapid initial decay of the electrical signal in (e). Beating between the two signals is evident around 430~ns in (e). Simulation parameters were $V_\mathrm{d}=8$~meV, $\fd=320$~MHz, $\fm=293$~MHz, $\Qm=700$, $\CCNTbar=0.1$~pF. To make the mechanical signal clear, an unrealistic $|\VGbar \frac{\partial \CCNT}{\partial u}| = 2.2$~aC/nm was chosen, approximately sixteen times larger than the geometrically expected value.
}
\end{figure*}

For a quantitative comparison with Fig.~2 we first extract numerical values of the circuit parameters by comparing the  reflection coefficient measured as in Fig.~1(b) with the result of a circuit model. Taking the known inductance $L=180$~nH and the fixed decoupling capacitance $\CD=87$~pF, and approximating $\CG \ll \CS$, three unknown parameters ($\CS$, $\CM$, $\Reff$) can be extracted from the width, depth, and center frequency of the measured resonance. We find that $\CS \approx 1.625$~pF, $\CM \approx 27$~pF, and $\Reff \approx 13~\Omega$ give a good match to the data~(Fig.~\ref{FigS2}).

For the experiments in Fig.~2, the drive was instead applied via the gate, so that $\VD=0$ and the driving voltage is 
\begin{equation}
\VG(t)=\VGbar + \delta \VG(t),
\end{equation}
where the time-independent part $\VGbar$ is much larger than the time-dependent part $\delta \VG(t)$. The capacitance 
\begin{equation}
\CCNT(t) = \CCNTbar + \delta \CCNT (t)
\end{equation} can likewise be separated into a time-independent part $\CCNTbar$ and a smaller time-dependent part $\delta \CCNT (t) = \partial \CCNT/ \partial u \cdot u(t)$. Then to lowest order the current driving the cavity is 
\begin{equation}
I_\mathrm{G}(t) = \CG \dot{\VG}(t) - \overline{V_\mathrm{G5}} \frac{\partial \CCNT}{\partial u} \dot{u}(t).
\label{eq:ICNT}
\end{equation}
Here the first term describes the direct capacitive coupling of the drive gate to the cavity circuit and the second term describes the mechanical part of the coupling. Henceforth we assume for simplicity that both DC gate voltages are equal, i.e. $\overline{V_\mathrm{G5}} = \VGbar$.

This current excites the cavity, giving rise to radiated voltage 
\begin{equation}
\Vout = I_\mathrm{G}\Ztrans,
\label{eq:Vout}
\end{equation}
where we have defined a circuit transimpedance
\begin{align}
\Ztrans	&\equiv \Vout/I_\mathrm{G}\\
		&= 	\frac{(Z_0||\ZM)\cdot(\ZS||(\Reff + Z_L + \ZD + (Z_0 || \ZM)))}{\Reff + Z_L + \ZD}
\label{eq:Ztrans}
\end{align}
with $\ZS, \ZD, \ZM$, and $Z_L$ respectively the impedance of $\CS, \CD, \CM$, and $L$, and $Z_0 \equiv 50~\Omega.$ This transimpedance is approximated by assuming a simple resonant cavity, giving:
\begin{equation}
\Ztrans \approx \Ztrans^\mathrm{max} \frac{if\fE/\QE}{f^2 - \fE^2 - i f \fE/\QE} e^{i\phiE},
\label{eq:Ztransapprox}
\end{equation}
where the effective cavity is parameterized by its resonance frequency $\fE$, quality factor $\QE$, maximum transimpedance $\Ztrans^\mathrm{max}$ and a phase offset $\phiE$. In our case, effective cavity parameters $\Ztrans^\mathrm{max} \approx 283~\Omega$, $\fE\approx 304$~MHz, and $\QE \approx 17.4$ provide a good approximation (Fig.~\ref{FigS2} inset).

\subsection{The vibrating nanotube}
We write the displacement in the driven mode as $U(x,t) = u(t) \xi(x)$, $u(t)$ is the time-dependent displacement parameter, $x$ is location along the nanotube and $\xi(x)$ is the mode eigenfunction~\cite{Poot2012}. The effective electromechanical force on the nanotube is 
\begin{equation}
F(t) = \frac{1}{2} \VG^2(t) \frac{\partial \CCNT}{\partial u}, 
\label{eq:force}
\end{equation}
whose time-dependent part is to lowest order~\cite{Sazonova2004}
\begin{equation}
\delta F (t) = \VGbar \frac{\partial \CCNT}{\partial u} \delta \VG(t).
\end{equation}
Transforming to the frequency domain, the corresponding displacement is~\cite{Aspelmeyer2014}
\begin{equation}
\delta u (f) = \chim(f) \delta F (f),
\label{eq:displacement}
\end{equation}
where the mechanical susceptibility is 
$\chim = \frac{1}{4\pi^2 m}[\fm^2 - f^2 + if\fm/\Qm]^{-1}$.		
Here $f$ is frequency, $m$ is the nanotube effective mass and $\fm$ and $\Qm$ are the mechanical resonance frequency and quality factor. We choose to normalize the mode eigenfunction such that $L^{-1}\int \xi^2(x) dx = 1$, where $L$ is the suspended length, thus making the effective mass equal to the mass of the suspended segment~\cite{Poot2012}.

\subsection{Response to pulsed driving}

To calculate the response to a pulsed drive as in Fig.~2, we transform Eq.~(\ref{eq:ICNT}) to the frequency domain and substitute into Eq.~(\ref{eq:Vout}). The resulting signal is
\begin{equation}
\Vout(f) = 2\pi i f\ \Ztrans(f)\ \CG\ (1 - \alpha \chim(f))\ \delta \VG (f),
\label{eq:response}
\end{equation}
where 
\begin{equation}
\alpha \equiv \frac{1}{\CG}\left(\VGbar \frac{\partial \CCNT}{\partial u}\right)^2
\end{equation}
is a parameter that characterizes the strength of the mechanical signal (mediated by the moving nanotube) relative to the electrical signal (due to direct excitation of the cavity).

\begin{figure}
\includegraphics[width=\columnwidth]{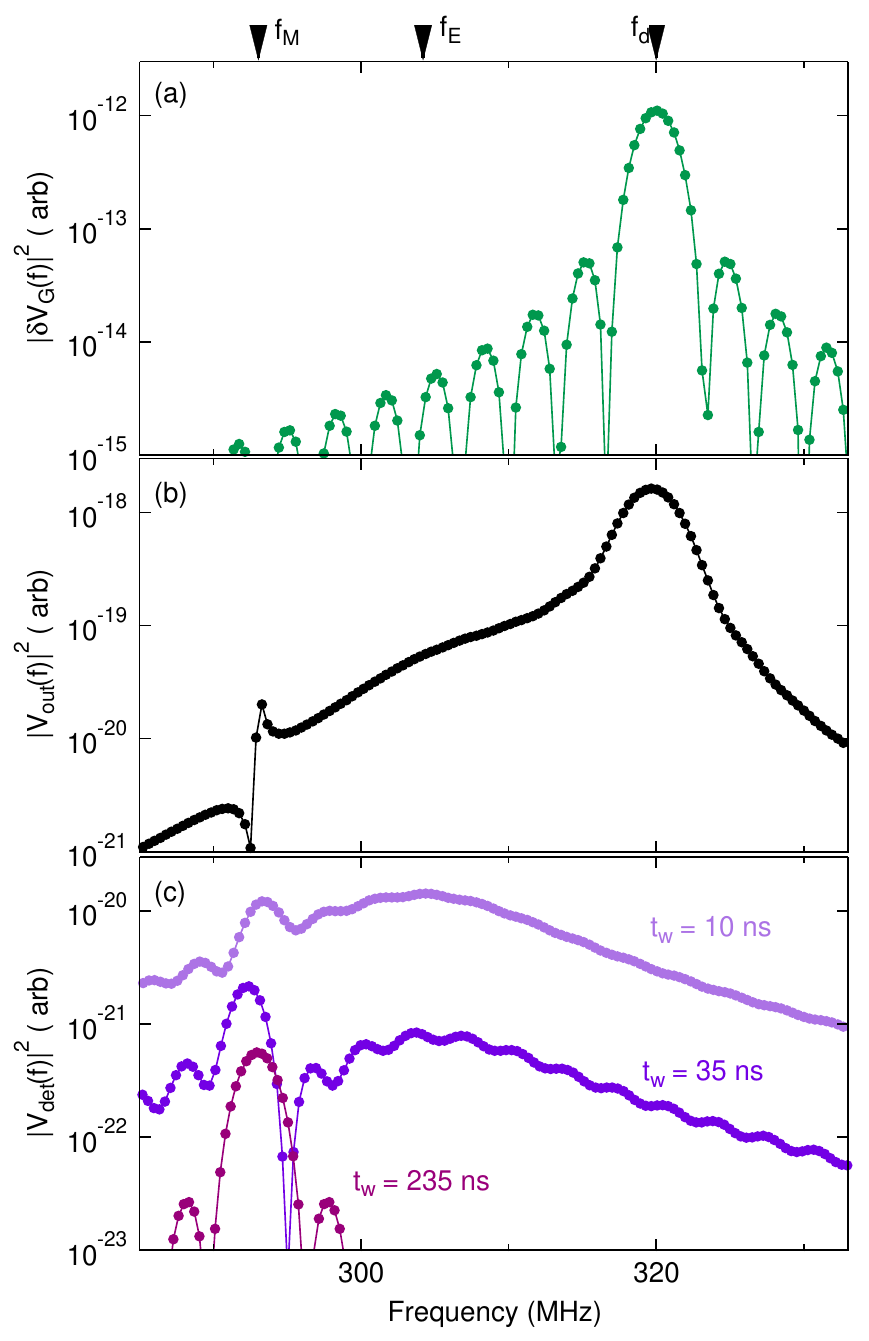}
\caption{\label{FigS4}
Power spectral densities calculated from simulations as in Fig.~\ref{FigS3}. (a) Drive signal (b) Cavity output (c) Gated output for three different delays $\tw$. As $\tw$ is increased, the slowly-decaying mechanical contribution becomes more prominent. Parameters are the same as in Fig.~\ref{FigS3}, except that a $2~\mu$s wait step has been added after the detection step to improve the frequency resolution of the power spectra. The three relevant frequencies are marked along the top axis.
}
\end{figure}

\begin{figure*}
\vspace{-0.3cm}
\includegraphics[width=17.8cm]{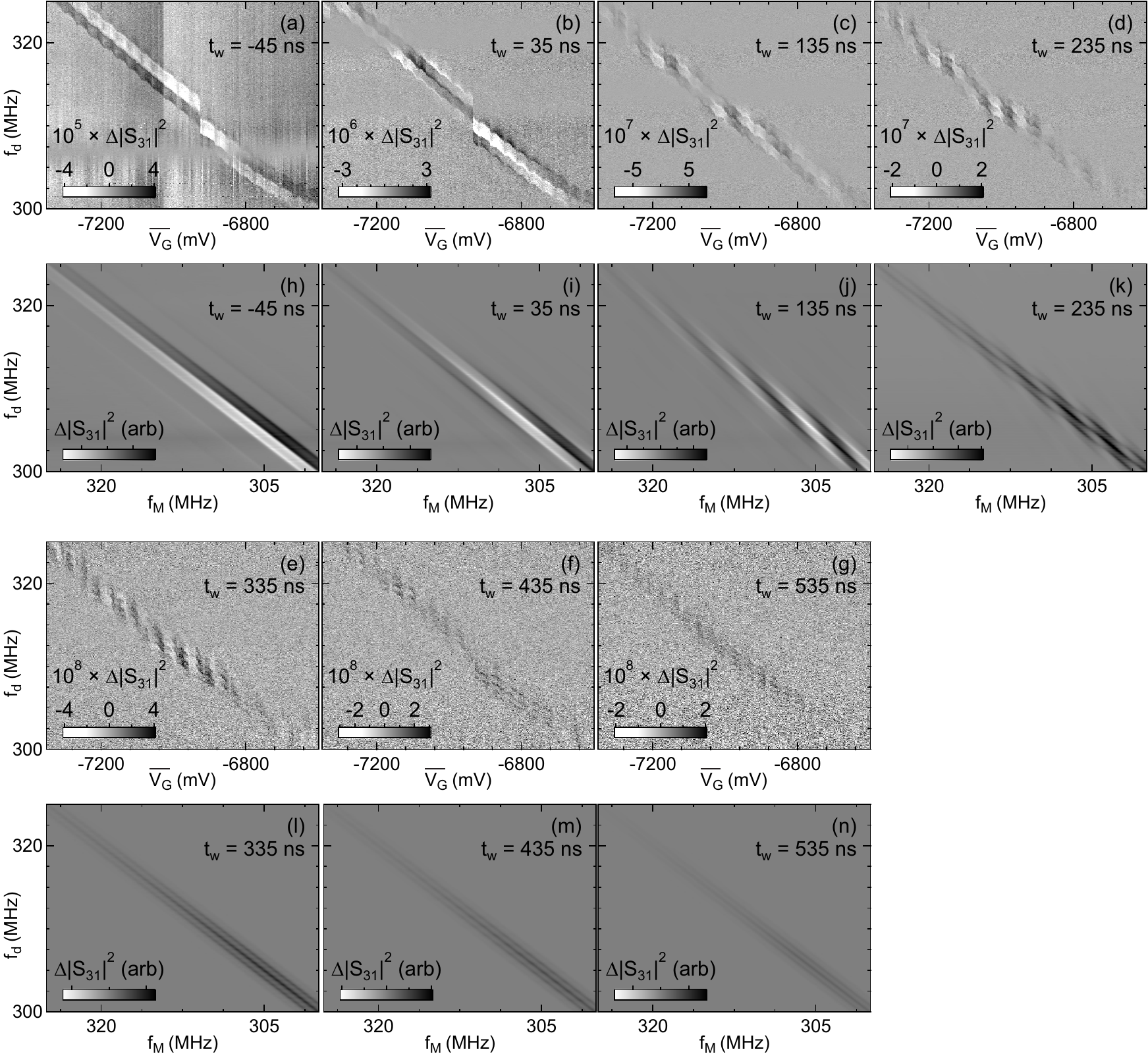}
\caption{\label{FigS5}
(a-e) measurements and (f-n) simulations of $\Delta \St^2$ for different $t_\mathrm{w}$. As in Fig.~2 of the main text, parameters are $\Qm=700$, $\alpha = 6 \times 10^{-11} \mathrm{Nm}^{-1}$. The drive burst and the detection window are both of 300~ns duration, and a wait of 70~ns is inserted between the end of the detection window and the beginning of the drive burst. To make the mechanical signal clearer in both data and simulation, the average of each row has been subtracted, and a separate color scale applied to each panel.
}
\end{figure*}

For the scheme of Fig.~2, typical signals involved are shown in Fig.~\ref{FigS3}. A gated drive signal $\delta \VG = G_\mathrm{in} (f) V_\mathrm{d} \cos (2\pi \fd t)$ (Fig.~\ref{FigS3}(c)), gives rise by Eq.~(\ref{eq:displacement}) to the displacement shown in Fig.~\ref{FigS3}(d) and by Eq.~(\ref{eq:response}) to the output signal of Fig.~\ref{FigS3}(e). The mechanical coupling is evident in the long tail of $\Vout$ during the undriven part of the cycle, which also shows beating between the electrical ringdown at frequency $\fE$ and the mechanical ringdown at frequency $\fm$. This mechanical signature is detected by gating the output, giving a detected signal 
\begin{equation}
\Vdet(t) = G_\mathrm{out}(t) \Vout(t).
\label{eq:Vdet}
\end{equation}
To obtain the simulated measurements of Fig.~2, $\Vdet(t)$ is simulated over 10 burst cycles using Eq.~(\ref{eq:Vdet}), and Fourier transformed using a Hanning window to give power spectra $|\Vdet(f)|^2$. To suppress numerical artifacts, the gating pulses are smoothed over $\sim 5$~ns.

Simulated power spectral densities are shown in Fig.~\ref{FigS4}. Whereas the drive and output signals are peaked at the drive frequency $\fd$, the detected signal contains a broad peak at $\fE$ from the electrical ringdown as well as a sharp peak at $\fM$ from the mechanical ringdown. With increasing $\tw$, the mechanical peak becomes dominant (Fig.~\ref{FigS4}(c)), allowing it to be isolated as in Fig.~4.

In the detection scheme of Fig.~2, the measured power is proportional to $|\Vdet(\fd)|^2$. A series of simulations for different $t_\mathrm{w}$ is shown in Fig.~\ref{FigS5}. The pattern can be understood as follows. The dominant signal occurs when the drive is resonant with the mechanical frequency, $\fd=\fM$. Beside this central peak, the modulation of the carrier gives rise to sidebands separated by multiples of the inverse burst period. The width of the central peak and sidebands is governed by the mechanical $Q$-factor. 

Superimposed on this pattern are a series of fringes arising from beating between the mechanical and cavity resonances, whose slope in a two-dimensional $(\fM, \fd)$ plot increases with increasing $\tw$. At short $\tw$, the sign of this beating is set by the phase of the mechanical displacement at the end of the drive burst, which (by Eq.~(\ref{eq:response})) is set by the phase of $\chim$; it is positive for $\fd<\fM$ and negative for $\fd>\fM$. Thus the fringes run nearly parallel to the $\fd=\fM$ line. However, after the end of the drive burst, the mechanical phase and the cavity phase advance at different frequencies ($\fM$ and $\fE$ respectively), building up a phase difference equal to $2\pi(\fM-\fE)\tw$. The effect of this phase difference is to tilt the fringe pattern as seen. At large $\tw$, the electrical signal becomes insignificant and the fringes disappear.

\subsection{Comparison of fit parameters with expectations}
\label{sec:numcomp}
The free parameters entering into the simulation using Eq.~(\ref{eq:response}) are (apart from an overall scale factor) $\alpha$ and $\Qm$. For the simulations in Fig.~2, these are chosen to yield a qualitatively similar fringe pattern and sideband sharpness as in the data, giving $\Qm \approx 700$ and $\alpha \approx 6 \times 10^{-11}$. The electromechanical signal is thus approximately $\alpha |\chim(\fm)| \approx 2 \times 10^{-6}$ times smaller than the direct electrical coupling between $\delta \VG$ and $\Vout$, confirming that a pulsed scheme is needed to detect it.

To compare these parameters with expectations, we assume a single-walled nanotube of diameter $D=4$~nm and length $L=650$~nm suspended a distance $a=110$~nm above the gates~\cite{Mavalankar2016}. These parameters lead to $m=6 \times 10^{-21}$~kg. Now using from the geometric capacitance
\begin{equation}
\frac{\partial \CCNT}{\partial u} \approx \frac{2\sqrt{2}\pi \epsilon_0 L_\mathrm{G}}{ a \left(\ln \frac{2a}{D}\right)^2},
\label{eq:dercap}
\end{equation}
where $L_\mathrm{G} = 50$~nm is the gate length, we obtain $\frac{\partial \CCNT}{\partial u} \approx 2.2 \times 10^{-12}~\mathrm{Fm}^{-1}$. With gate voltage $\VGbar = 7$~V, assuming a geometric capacitance between gates 4 and 5 of $\CG \approx 0.1$~pF gives $\alpha \approx 2.4 \times 10^{-9}~\mathrm{Nm}^{-1}$. This estimate ignores the tunnel barriers between the nanotube quantum dot and the leads, which by acting as a series impedance reduce the effective coupling and therefore reduce $\alpha$. It also ignores the fact that in this device both driving and coupling gates are offset from the nanotube centre, which reduces the coupling, and neglects fringing capacitance, which increases it.

\subsection{Parasitic electrical resonances}
\begin{figure}
	\includegraphics[width=\columnwidth]{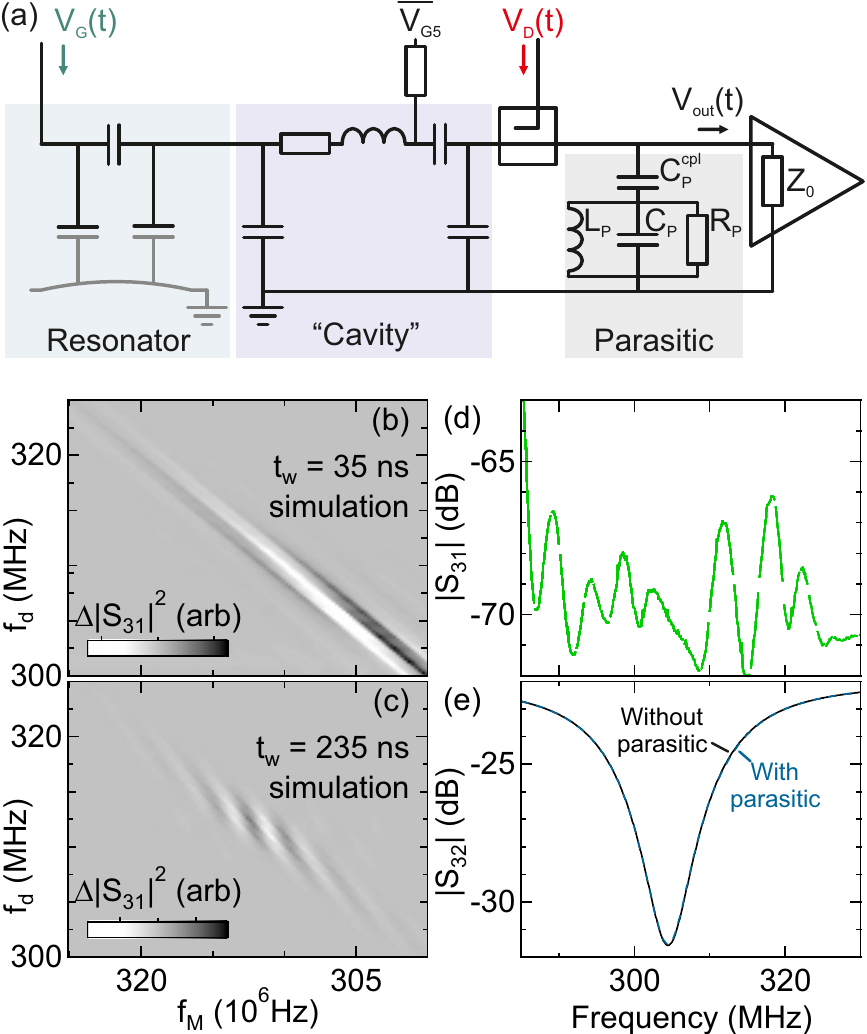}
	\caption{\label{FigS6b}
		(a) Circuit of Fig.~\ref{FigS1} including a weakly coupled parasitic mode, modeled as an LCR resonator. (b-c) Simulated $\Delta|S_{31}|^2$ in this model. To approximately match the data, chosen parameters are $\LP=250$~nH, $\CP=1$~pF, $\RP=30~\mathrm{k}\Omega$, and $\CPcpl=0.04$~pF, giving parasitic resonance frequency $\fP\approx 312$~MHz and quality factor $\QP\approx 60$.  As seen, the strongest signal in panel (c) shifts towards the parasitic frequency. (d) Measured $|S_{31}|$ background signal for data set of Fig.~2(c). This signal is of purely electrical origin, and its peaks confirm the presence of weakly coupled parasitic resonances in the cryostat. (e) Simulated tank circuit reflectance in continuous-wave configuration with and without the parasitic. The parasitic is barely distinguishable in this configuration.
	}
\end{figure}

As seen especially in Fig.~2(c) in the main text, the strongest mechanical signal is not always at the exact cavity frequency. This is because at long $\tw$ the signal is affected by weakly coupled parasitic resonances of the cryostat. To show this, we perform the same simulation as above, with the circuit of Fig.~\ref{FigS1} modified by addition of a weakly coupled parasitic resonance~(Fig.~\ref{FigS6b}(a)). With appropriate parameters, this gives a stronger signal at the parasitic frequency ($\sim 312$~MHz) than at the cavity frequency $\fE \approx 304$~MHz~(Fig.~\ref{FigS6b}(b-c)). Although such a parasitic mode barely affects the circuit scattering parameters in a continuous-wave measurement as in Fig.~1, it is evidenced by a background frequency dependence of $|S_{31}|$, independent of the mechanics, in a pulsed measurement~(Fig.~\ref{FigS6b}(d)). Our simulation does not allow the large number of parasitic circuit parameters to be extracted, but does show quantitatively that parasitics can explain the observed effect. Since only modes with quality factors higher than the cavity give such an effect, it could be eliminated by modest improvements to the cavity.

\subsection{Nonlinear effects}

\begin{figure}
	\includegraphics[width=\columnwidth]{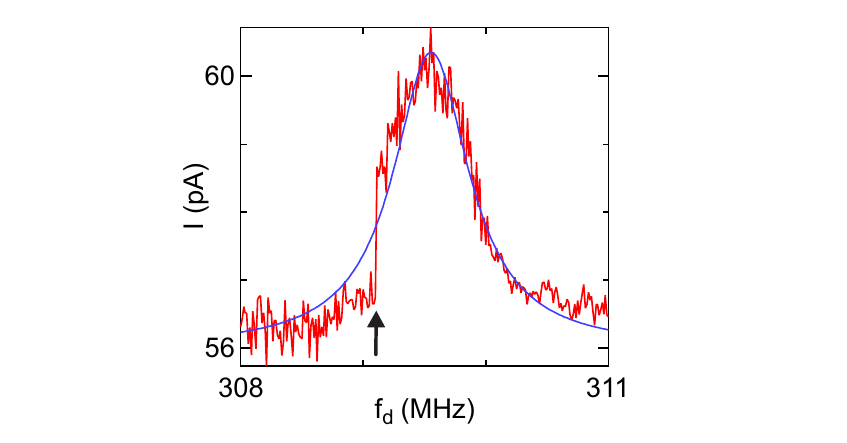}
	\caption{\label{FigS6}
		Red: current as a function of $\fd$ showing a peak at mechanical resonance. The peak is well fit by a Lorentzian (blue) as expected for linear response. However, a reproducible switch (arrow) indicates a bistability caused by a nonlinear contribution to the restoring force. For this data, $\VGbar=-6930$~mV, $\Vsd=10$~mV and the drive power is $P=-38$~dBm.}
\end{figure}

This simulation makes two assumptions of linearity. First, Eq.~(\ref{eq:displacement}) assumes that the resonator experiences a linear restoring force. This can be tested by inspecting the resonant lineshape. Whereas a linear resonator should display a Lorentzian response, nonlinearity is indicated by asymmetry and bistability~\cite{Steele2009}. A typical response curve under continuous driving is shown in Fig.~\ref{FigS6}. Although a switch indicating bistability is seen, the lineshape is predominantly Lorentzian, indicating that the linear restoring force dominates. Of course, as the resonator rings down non-linear effects should become weaker, justifying the approximation.

Second, Eq.~(\ref{eq:ICNT}) assumes linearity of the motion transduction. To verify this, we have estimate the shift in $\omega_E$ corresponding to the maximum possible displacement of the nanotube. This upper bound is given by the distance between the equilibrium position of the nanotube and the gate electrodes, which is approximately 110~nm. For parameters detailed in the previous section, this shift is approximately 22~Hz, orders of magnitude smaller than the cavity linewidth, which is of order 10~MHz. The transduction regime is therefore linear for our setup.  

\section{Numerical estimates of optomechanical parameters}

\subsection{Vacuum optomechanical coupling}

The zero-point motion of the CNT is 
\begin{equation}
\uzp=\sqrt{\frac{\hbar}{4\pi \ensuremath{m}\fm}} \approx 2~\mathrm{pm}
\end{equation}
for typical frequency $\fm=310$~MHz. Using this value of $\uzp$, taking $\frac{\partial \CCNT}{\partial u}$ from Eq.~(\ref{eq:dercap}), and estimating $\CS\approx1.625$~pF from the circuit simulation, Eq.~(\ref{eq:g0}) gives $\go\approx2\pi\times0.4$ mHz.

\subsection{Sensitivity}

The amplitude sensitivity $\Su$ is related to the amplifier voltage sensitivity $\Sv$ by 
\begin{equation}
\Su=\frac{\Sv}{\partial \Vout/\partial u}.
\label{eq:Su}
\end{equation}
Here $\Sv=\sqrt{\kB\TN\ZO}$ is the voltage noise spectrum, with $\kB$ the Boltzmann constant and $\TN$ the amplifier noise temperature. From Eqs.~(\ref{eq:ICNT}-\ref{eq:Vout}) and considering $u(t)$ harmonic, we have
\begin{equation}
 \frac{\partial \Vout}{\partial u}=2\pi\fm\VGbar \frac{\partial \CCNT}{\partial u} \Ztrans.
\label{eq:der}
\end{equation}
Taking the experimentally known gate voltage  $\VGbar = 7$~V, the specified noise temperature $\TN=3.7$~K, and using $\Ztrans\approx283~\Omega$ from the circuit simulation (Sec.~\ref{sec:circuit}), we estimate $\Su=6$~pm$/\sqrt{\text{Hz}}$ in the measured setup.

\subsection{Approach to the quantum limit}

The standard quantum limit of sensitivity corresponds approximately to resolving the zero-point fluctuations within the resonator lifetime. More specifically, it requires sensitivity such that the imprecision $\Delta u = \Su \sqrt{\Delta \fM}$ approaches $\Delta \uQL$, where $\Delta \fM=2\pi\fm/(4\Qm)$ is the effective noise bandwidth of the mechanical oscillator and $\Delta \uQL = \sqrt{2/\ln3}\,\uzp$ is the ultimate resolution imposed by the uncertainty principle~\cite{LaHaye2004}.

For our scheme,  the mechanical linewidth is $\Delta \fM \approx 700$~kHz and therefore $\Delta u \approx 1700 \times \Delta \uQL$. 
To approach the quantum limit, we would need a quantum-limited amplifier, with a noise temperature~\cite{Caves1982}

\begin{equation}
\TN^\mathrm{min}=\frac{h \fM}{\ln 3~k_B} \approx 10~\mathrm{mK}.
\end{equation}
In an optimized device with $L_\mathrm{G} = 550$~nm, Eq.~(\ref{eq:dercap}) predicts an increase of the coupling capacitance to $\frac{\partial \CCNT}{\partial u}\approx 2.4 \times 10^{-11}~\mathrm{Fm}^{-1}$. If $\VGbar$ could then be increased to 65~V, Eq.~(\ref{eq:Su}) predicts $\Su=3.5$~fm$/\sqrt{\text{Hz}}$. We thus obtain an amplitude imprecision $\Delta u\approx 3~\mathrm{pm} \approx \Delta \uQL$. The transduction regime remains linear in this setup configuration, since $\omega_E$ still shifts by only 250~Hz for full nanotube displacement.

\vfill